\documentclass[aps,prd,final,nofootinbib]{revtex4}
\pdfoutput=1
\usepackage{amssymb,amsmath,amstext,amsfonts}
\usepackage{bbold,ulem,bm}
\usepackage{graphics}
\usepackage{mathtools}
\usepackage{hyperref}
\usepackage{color}

\usepackage{graphicx}
\usepackage{mwe}
\usepackage{mathrsfs}

\newcommand{\beq}{\begin{equation}}
\newcommand{\eeq}{\end{equation}}
\newcommand{\ba}{\begin{eqnarray}}
\newcommand{\ea}{\end{eqnarray}}
\newcommand{\bal}{\begin{aligned}}
\newcommand{\eal}{\end{aligned}}
\newcommand{\Lag}{{\mathcal{L}}}

\begin{document}

\title{On the Coupling of Generalized Proca Fields \\
to Degenerate Scalar-Tensor Theories}

\author{Sebastian Garcia-Saenz}
\affiliation{Theoretical Physics, Blackett Laboratory, Imperial College London, SW7 2AZ London, U.K.}

\begin{abstract}
We prove that vector fields described by the generalized Proca class of theories do not admit a consistent coupling to a gravitational sector defined by a scalar-tensor theory of the degenerate type. Under the assumption that there exists a frame in which the Proca field interacts with gravity only through the metric tensor, our analysis shows that at least one of the constraints associated with the degeneracy of the scalar-tensor sector is inevitably lost whenever the vector theory includes a coupling to the Christoffel connection.
\end{abstract}

%\pacs{}
%\keywords{}

\maketitle

%%%%%%%%%%%%%%%%%%%%%%%%%%%%%%%%%%%%%%%%%%%
%%%%%%%%%%%%%%%%%%%%%%%%%%%%%%%%%%%%%%%%%%%

\section{Introduction}

The extension of general relativity (GR) by additional light degrees of freedom is arguably the most natural way to provide a dynamical explanation of dark energy, thereby dispensing of the cosmological constant as the source of the observed late-time cosmic acceleration. Considering a single scalar field in addition to the metric tensor is, in this regard, particularly well motivated. These so-called scalar-tensor theories of gravity \cite{Fujii:2003pa,Quiros:2019ktw} thus provide the most minimal modification of Einstein gravity in terms of local degrees of freedom and under some standard assumptions such as Poincar\'e invariance and locality. This is a virtue both from the theoretical and experimental perspectives, as its relative simplicity allows for strong analytical control while maintaining much of the phenomenology of GR. It is also not the least telling case for scalar-tensor theories that the related mechanism of inflation was likely to be at work during the pre-Big Bang epoch.\footnote{In the context of higher-derivative theories, it is worth remarking that the Starobinsky model of inflation provides a healthy and consistent example of this class \cite{Starobinsky:1979ty}. See also \cite{Starobinsky:2016kua,Mironov:2019qjt,Arbuzov:2020pgp} for related works on inflationary models based on generalized scalar-tensor theories.}

The complete classification of scalar-tensor theories thus seems to be an interesting and timely theoretical problem. In this effort, the assumption of having precisely three local degrees of freedom --- two propagated by the metric and one by the scalar field --- severely restricts the space of possible models. Although the physically meaningful question should make a distinction of light versus heavy degrees of freedom, it has nevertheless proved fruitful to demand the strict absence of additional fields beyond the aforementioned three, seeing that the resulting models often enjoy interesting properties that may have been difficult to discover through a more agnostic construction based on the rules of effective field theory.

This restriction on the number of degrees of freedom makes the classification problem mathematically well defined, although not easy as it turns out. Given the symmetries of the theory, it is sufficient to demand second order field equations, and taking this as a premise the problem has indeed been fully solved. The solution is given by Horndeski's scalar-tensor theory \cite{Horndeski:1974wa,Nicolis:2008in,Deffayet:2009wt,Deffayet:2009mn,Deffayet:2011gz}. The remarkable observation is that this premise is however not a necessary one. That is, higher order equations of motion are not necessarily associated to extra unwanted degrees of freedom --- unwanted indeed as they are generically associated to ghost-type instabilities according to the Ostrogradski theorem. This is so because the equations may happen to be degenerate, in the sense that a subset of them follows as a consequence of the others, implying in particular a reduction of the number of pieces of initial data that one would have naively inferred. The development and classification of these so-called degenerate scalar-tensor theories has been an active research program over the past decade \cite{Zumalacarregui:2013pma,Lin:2014jga,Gleyzes:2014dya,Gleyzes:2014qga,Langlois:2015cwa,Crisostomi:2016tcp,Crisostomi:2016czh,Achour:2016rkg,BenAchour:2016fzp}. New models have been discovered throughout the years and have been given different names. We will refer to all of them collectively as DHOST, an acronym that stands for ``Degenerate Higher-Order Scalar-Tensor" theories. See \cite{Deffayet:2013lga,Langlois:2018dxi,Kobayashi:2019hrl} for reviews.

DHOST theories provide then a very interesting solution to the classification problem of scalar-tensor gravity. They are consistent theories within the scope of that problem, at least according to the way we have formulated it, although it is clear that physical consistency will reduce the space of allowed models by the imposition of further constraints. Most of these constraints arise from experimental tests of gravity, although here we will not be concerned with them --- not because they are not important, but because their importance is contingent on the physical context. For instance constraints derived from cosmological observations \cite{DeFelice:2015isa,Sakstein:2016ggl,Creminelli:2017sry,Peirone:2017ywi,Dima:2017pwp,Creminelli:2018xsv,Creminelli:2019kjy,Hirano:2019scf,Creminelli:2019nok,Noller:2020afd,Anson:2020fum} need not apply on the scales of compact astrophysical objects. Theoretical constraints on the other hand have the chance to be more generally applicable, even if experiments must have the last word.

One such theoretical constraint that has remained largely overlooked is the question on the consistency of matter coupling in DHOST theories. The fact that matter fields can be problematic is seen easily in the Hamiltonian language, in which the degeneracy of the field equations is manifested in the form of a constraint on the phase space variables. The mixing with matter fields can then obstruct this constraint, leading to the reappearance of the ghost degree of freedom and an inconsistent theory \cite{deRham:2016wji}. This may occur even if matter is minimally coupled to the metric tensor, for an indirect coupling with the DHOST scalar is still present. It is worth remarking that this issue is of course not specific to DHOST theories and may happen whenever two theories, where either or both have constraints when considered separately, are coupled in some way \cite{Isenberg:1977fs}. It is thus a virtue of the Hamiltonian language to make it manifest that the degeneracy condition is in truth a constraint, on equal footing to other constraints.

Understanding the precise ways in which the DHOST constraint may be lost was the subject of the work \cite{Deffayet:2020ypa}. Let us denote the constraint by $\Psi\approx 0$, where $\Psi$ is a phase space function to be made explicit later, and the symbol ``$\approx$'' means weak equality. We can then distinguish two types of pathological matter theories:
\begin{itemize}
\item[(I)] The constraint $\Psi$ is lost, and no analogue of it exists.

\smallskip

This will be the case when the rank of the Hessian matrix
\beq
\mathcal{H}_{IJ}:=\frac{\partial^2\mathcal{L}}{\partial \dot{\psi}^I\partial \dot{\psi}^J}\,,
\eeq
(here $\psi^I$ stands for all the fields) is greater than the sum of the ranks of the DHOST and matter Hessians that one would have in the absence of coupling. This cannot occur when the full Hessian is block-diagonal in the DHOST and matter variables. As we are restricting our attention to minimal matter coupling, any matter Lagrangian that does not involve the Christoffel connection will lead to a block-diagonal Hessian and thus be safe according to this criterion. The converse of this is of course not true. Although a non-block-diagonal Hessian is at risk of failing this consistency check, it may still enjoy a (possibly modified) degeneracy constraint.

\medskip

\item[(II)] The constraint $\Psi$ (or some analogue of it) does exist, but it fails to Poisson-commute with one or more constraints present in the matter sector.

\smallskip

In the absence of matter the DHOST constraint $\Psi$ is a primary, second-class constraint, and it Poisson-commutes with all the other primary constraints in the gravity sector. It therefore leads to a secondary constraint, which together with $\Psi$ is responsible for removing the would-be ghost degree of freedom. If now the matter sector itself has some constraints, there is the risk that they may not commute with $\Psi$, implying the loss of the associated secondary constraint and the reappearance of the unwanted degree of freedom.

\end{itemize}

\medskip

It is not difficult to find examples that fail either of these two criteria; some explicit pathological matter models were studied in \cite{Deffayet:2020ypa}. The aim of the present article is to analyze these consistency criteria in detail for a more interesting model, namely the generalization of the Proca theory of a massive spin-1 field \cite{Tasinato:2014eka,Heisenberg:2014rta,Hull:2015uwa,Jimenez:2016isa}. This class of models, dubbed Generalized Proca (GP), has been subject to intense scrutiny for its potential role in cosmology as a dark energy fluid and also in the physics of compact astrophysical objects \cite{DeFelice:2016yws,DeFelice:2016uil,Minamitsuji:2016ydr,Babichev:2017rti,Heisenberg:2017xda,Kase:2018owh,Kase:2020yhw,Garcia-Saenz:2021uyv}. GP theory extends the linear Proca model by the inclusion of derivative interactions while maintaining the constraints that ensure that one of the components of the vector field is non-dynamical. The theory thus falls into the ``dangerous'' class of matter fields when coupled to DHOST: the non-trivial interactions produce a coupling to the Christoffel connection upon covariantization, while the Proca constraint risks spoiling the Poisson algebra of the coupled DHOST-GP system.

Our main result is the proof that GP theory cannot be consistently coupled to DHOST gravity within the framework we consider. The main assumptions are the following: (i) we focus exclusively on the so-called quadratic DHOST class, i.e.\ scalar-tensor theories whose Lagrangian involves operators that are at most quadratic in $\nabla^2\phi$ (here $\phi$ is the scalar field); (ii) we consider a truncated version of GP theory with at most cubic derivative self-interactions; (iii) the GP vector field couples to the DHOST sector only through the metric tensor. Assumptions (i) and (ii) are not essential and we expect all our results to hold for more general DHOST models as well as for the complete GP Lagrangian. Assumption (iii) is on the other hand more restrictive, but is certainly reasonable and in line with our set-up of treating the Proca field as a matter field which couples to gravity in accordance with the equivalence principle. We will come back to this point in the final discussion.

%%%%%%%%%%%%%%%%%%%%%%%%%%%%%%%%%%%%%%%%%%
\section{ADM decomposition of DHOST and GP theories}

In this section we review the definitions of the DHOST and GP theories that we focus on in this article. We then perform a $3+1$ decomposition of the Lagrangians in terms of ADM variables.

\subsection{DHOST Lagrangian}

The gravitational sector of our framework is given by the quadratic DHOST Lagrangian,
\beq \label{eq:full quad dhost}
S_{\rm g}[\phi,g]=\int d^4x\sqrt{-g}\Big[F(\phi,X)R+P(\phi,X)+Q(\phi,X)\Box\phi+C^{\mu\nu\rho\sigma}[\phi]\nabla_{\mu}\nabla_{\nu}\phi\nabla_{\rho}\nabla_{\sigma}\phi\Big] \,.
\eeq
Here $R$ is the curvature scalar constructed from the metric $g_{\mu\nu}$, while $F$, $P$ and $Q$ are generic functions of the scalar field $\phi$ and
\beq
X:=\nabla^{\mu}\phi\nabla_{\mu}\phi \,.
\eeq
The tensor $C^{\mu\nu\rho\sigma}$ is defined as
\beq\bal
C^{\mu\nu\rho\sigma}&:=A_1g^{\mu(\rho}g^{\sigma)\nu}+A_2g^{\mu\nu}g^{\rho\sigma}+\frac{A_3}{2}\left(\phi^{\mu}\phi^{\nu}g^{\rho\sigma}+\phi^{\rho}\phi^{\sigma}g^{\mu\nu}\right) \\
&\quad+\frac{A_4}{2}\left(\phi^{\mu}\phi^{(\rho}g^{\sigma)\nu}+\phi^{\nu}\phi^{(\rho}g^{\sigma)\mu}\right)+A_5\phi^{\mu}\phi^{\nu}\phi^{\rho}\phi^{\sigma} \,,
\eal\eeq
where $\phi_{\mu}:=\nabla_{\mu}\phi$ and the $A$'s are also functions of $\phi$ and $X$.

For the purpose of analyzing the constraints in the Hamiltonian language we carry out a time-space split or $3+1$ decomposition of the Lagrangian. The metric tensor is expanded in ADM variables \cite{Arnowitt:1962hi}, i.e.\ the lapse $N$, shift $N^i$ and 3-metric $\gamma_{ij}$,
\beq
g_{\mu\nu}=\left(\begin{array}{cc}
-N^2+N^kN_k & N_i \\
N_j & \gamma_{ij} \end{array}\right) \,,\qquad g^{\mu\nu}=\frac{1}{N^2}\left(\begin{array}{cc}
-1 & N^i \\
N^j & N^2\gamma^{ij}-N^iN^j \end{array}\right) \,,
\eeq
and the measure factor is $\sqrt{-g}=N\sqrt{\gamma}$. Spatial indices are raised and lowered with the 3-metric and its inverse, so for example $N_i=\gamma_{ij}N^j$ (the shift function is defined with an upper index). The extrinsic curvature of the constant-time hypersurfaces is
\beq
K_{ij}=\frac{1}{2N}\left(\dot{\gamma}_{ij}-2D_{(i}N_{j)}\right) \,,
\eeq
where $D_i$ is the covariant derivative compatible with the 3-metric and a dot denotes differentiation with respect to the time coordinate $x^0=t$. We also introduce
\beq
n^{\mu}:=\frac{1}{N}\left(1,-N^i\right)\,,\qquad a^{\mu}:=n^{\nu}\nabla_{\nu}n^{\mu}=\left(0,\frac{D^iN}{N}\right) \,,
\eeq
and note that $n^{\mu}n_{\mu}=-1$.

In the Hamiltonian language one introduces a canonical momentum associated to each field velocity. The DHOST Lagrangian is a function of the second derivative of the scalar field, therefore both $\phi$ and $\nabla_{\mu}\phi$ have conjugate momenta in phase space. It is convenient to introduce an auxiliary vector field $A_{\mu}$ which is constrained as $A_{\mu}=\nabla_{\mu}\phi$ by means of a Lagrange multiplier \cite{Deffayet:2015qwa,Langlois:2015cwa,Langlois:2015skt}. Thus the modified DHOST action we will inspect is
\beq\bal
S_{\rm g}[\phi,A,g]&=\int d^4x\sqrt{-g}\Big[F(\phi,X)R+P(\phi,X)+Q(\phi,X)\nabla_{\mu}A^{\mu} \\
&\quad +C^{\mu\nu\rho\sigma}[\phi,A]\nabla_{\mu}A_{\nu}\nabla_{\rho}A_{\sigma}+\lambda^{\mu}(A_{\mu}-\nabla_{\mu}\phi)\Big] \,,
\eal\eeq
where it is understood that every instance of $\nabla_{\mu}\phi$ in $C^{\mu\nu\rho\sigma}$ has been replaced by $A_{\mu}$, and similarly $X$ now stands for $A^{\mu}A_{\mu}$. The Lagrangian is now purely first order in derivatives and the passage to the Hamiltonian proceeds as usual. Following the analysis of \cite{Langlois:2018dxi} we decompose the vector $A_{\mu}$ in its spatial components $A_i$ and the redefined time component
\beq
A_{*}:=n^{\mu}A_{\mu}=\frac{1}{N}\left(A_0-N^iA_i\right) \,.
\eeq
Details of the $3+1$ decomposition may be found in \cite{Langlois:2018dxi,Deffayet:2020ypa} so here we only quote the final result:
\beq\bal \label{eq:full adm dhost}
S_{\rm g}&=\int dtd^3x\Big\{N\sqrt{\gamma}\Big[{\cal A}V_{*}^2+2{\cal B}^{ij}V_{*}K_{ij}+{\cal K}^{ij,kl}K_{ij}K_{kl}+2{\cal C}^{ij}K_{ij}+2{\cal C}^0V_{*}-{\cal U}\Big]\\
&\quad+\lambda^0\left(NA_{*}+N^iA_i-\dot{\phi}\right)+\lambda^i\left(A_i-D_i\phi\right)\Big\} \,,
\eal\eeq
where
\beq
V_{*}:=\frac{1}{N}\left(\dot{A}_{*}-\Xi_A\right) \,,\qquad \Xi_A:=A^iD_iN+N^iD_iA_{*} \,.
\eeq
The coefficients appearing in \eqref{eq:full adm dhost} are given explicitly as follows:
\beq\bal
{\cal A}&=A_1+A_2-(A_3+A_4)A_{*}^2+A_5A_{*}^4 \,, \\
{\cal B}^{ij}&=A_{*}\left(2F_X+A_2-\frac{A_3}{2}\,A_{*}^2\right)\gamma^{ij}-\frac{A_{*}}{2}\left(A_3+2A_4-2A_5A_{*}^2\right)A^iA^j \,, \\
{\cal K}^{ij,kl}&=\left(F+A_1A_{*}^2\right)\gamma^{i(k}\gamma^{l)j}-\left(F-A_2A_{*}^2\right)\gamma^{ij}\gamma^{kl} +\left(2F_X-\frac{A_3}{2}\,A_{*}^2\right)\left(\gamma^{ij}A^kA^l+\gamma^{kl}A^iA^j\right) \\
&\quad -A_2\left(A^iA^{(k}\gamma^{l)j}+A^jA^{(k}\gamma^{l)i}\right)-\left(A_4-A_5A_{*}^2\right)A^iA^jA^kA^l \,, \\
\eal\eeq
while the expressions for ${\cal C}^{ij}$, ${\cal C}^0$ and ${\cal U}$ (which multiply terms that are at most linear in the velocities) will not be needed in our analysis; the interested reader may find them in Appendix \ref{sec:appA}.

The degeneracy of the DHOST Lagrangian is manifested in the fact that the determinant of the Hessian matrix of second time derivatives vanishes identically,\footnote{It is important to remark that the vanishing of the Hessian determinant implies the existence of a primary constraint, which is not by itself enough to remove a full Lagrangian degree of freedom. Nevertheless, the existence of an associated secondary constraint is guaranteed by the general covariance of the action, as shown explicitly in \cite{Langlois:2015cwa}. The same consideration holds for the situation when the DHOST constraint is modified in the presence of matter (as exemplified in eq.\ \eqref{eq:L4 modified dhost cond}), provided of course the latter admits a consistent coupling according to the criteria explained in the introduction.}
\beq \label{eq:dhost cond}
{\cal A}-{\cal K}^{-1}_{ij,kl}{\cal B}^{ij}{\cal B}^{kl}=0 \,.
\eeq
This relation translates into a set of algebraic equations for the coefficient functions $A_I$, $I=1,2,3,4,5$, and the solutions have been classified in \cite{Langlois:2015cwa}. Note the implicit assumption that the gravitational kinetic matrix ${\cal K}$ must be invertible, ensuring that DHOST can be connected smoothly, in theory space, to standard GR. The inversion of ${\cal K}$ can be done explicitly and the reader may find the result in Appendix \ref{sec:appB}.

\subsection{GP Lagrangian}

GP is a vector-tensor theory that describes the coupled dynamics of a vector field $B_{\mu}$ and metric $g_{\mu\nu}$. In isolation, this theory is consistent in the sense that it describes $3+2$ degrees of freedom, corresponding to massive spin-1 and massless spin-2 particles, at the complete non-linear level. The Lagrangian is given by \cite{Tasinato:2014eka,Heisenberg:2014rta}
\beq
S_{\rm GP}=\int d^4x\sqrt{-g}\sum_{I=2}^5 \Lag_I[g,B]\,,
\eeq
and we have explicitly
\beq\bal \label{eq:GP terms def}
\Lag_2&=G_2(Y,{\cal F},{\cal G}) \,, \\
\Lag_3&=G_3(Y)\nabla_{\mu}B^{\mu} \,, \\
\Lag_4&=G_4(Y)R-2G_{4}'(Y)\Big[(\nabla_{\mu}B^{\mu})^2-\nabla_{\mu}B^{\nu}\nabla_{\nu}B^{\mu}\Big] \,, \\
\Lag_5&=G_5(Y)G^{\mu\nu}\nabla_{\mu}B_{\nu} +\frac{G_5(Y)}{3}\Big[(\nabla_{\mu}B^{\mu})^3-3(\nabla_{\rho}B^{\rho})\nabla_{\mu}B^{\nu}\nabla_{\nu}B^{\mu}+2\nabla_{\mu}B^{\nu}\nabla_{\nu}B^{\rho}\nabla_{\rho}B^{\mu}\Big] \,, \\
\eal\eeq
with the definitions
\beq
Y:=B^{\mu}B_{\mu}\,,\qquad {\cal F}:=-\frac{1}{4}\,B^{\mu\nu}B_{\mu\nu} \,,\qquad {\cal G}:=B^{\mu}B^{\nu}B_{\mu}^{\phantom{\mu}\rho}B_{\nu\rho} \,,
\eeq
and $B_{\mu\nu}:=\nabla_{\mu}B_{\nu}-\nabla_{\nu}B_{\mu}$, while $R$ and $G_{\mu\nu}$ are respectively the curvature scalar and Einstein tensor constructed from the metric $g_{\mu\nu}$. A prime on the coefficient functions denotes differentiation with respect to the argument $Y$, e.g.\ $G_4'\equiv\frac{dG_4}{dY}$. The operators in \eqref{eq:GP terms def} do not exhaust the whole GP class. We do not expect the additional terms to affect any of our conclusions, so the truncated model we consider is general enough to illustrate the message of this paper. See the final discussion section for further comments on this point.

Like DHOST, GP is a degenerate theory in the sense that not all among the components of $B_{\mu}$ are dynamical. As is well known, in the standard Proca theory there exists a (local) frame in which $B_0$ does not propagate, and GP theory is precisely constructed so as to generalize this property to include non-trivial derivative interactions. In the Hamiltonian language, this degeneracy will manifest itself in the fact that the kinetic part of the Lagrangian (i.e.\ the operators that are at least quadratic in the velocity variables) will be independent of the time component of the vector field velocity.

In the following subsections we detail the $3+1$ decomposition of the operators entering in the GP terms defined above. The metric is again expanded in ADM variables while the Proca field, similarly to the DHOST auxiliary vector $A_{\mu}$, is decomposed in its spatial part $B_i$ and
\beq
B_{*}:=n^{\mu}B_{\mu}=\frac{1}{N}\left(B_0-N^iB_i\right) \,.
\eeq
The reader not interested in the particulars may skip to the next section where we provide the relevant collected results.

\subsubsection{$\Lag_2$ term}

The GP term $\Lag_2$ is a generic function of the scalars $Y$, ${\cal F}$ and ${\cal G}$. Expanding in ADM components we find
\beq\bal
Y&=-B_{*}^2+B^iB_i\,,\\
{\cal F}&=\frac{1}{2}\,F^iF_i-\frac{1}{4}\,b^{ij}b_{ij}\,,\\
{\cal G}&=B_{*}^2F^iF_i-B^iB^jF_iF_j-2B_{*}B^i\left(F^jb_{ij}+\frac{N^j}{N}\,b_i^{\phantom{i}k}b_{jk}\right)-\frac{N^iN^j}{N^2}\,B_{*}^2b_i^{\phantom{i}k}b_{jk}\,,
\eal\eeq
where
\beq\bal \label{eq:def F_i}
F_i&:=\frac{1}{N}\left(\dot{B}_i-\Upsilon_i\right)\,,\qquad \Upsilon_i:=D_i(NB_{*})+B_jD_iN^j+N^jD_jB_i\,,\\
b_{ij}&:=D_iB_j-D_jB_i\,.
\eal\eeq
Therefore $\Lag_2$ is manifestly degenerate as it is independent of $\dot{B}_{*}$.

\subsubsection{$\Lag_3$ term}

For the GP term $\Lag_3$ we only need the expression
\beq
\nabla_{\mu}B^{\mu}=-W_{*}-B_{*}K+D_iB^i \,,
\eeq
where
\beq
W_{*}:=\frac{1}{N}\left(\dot{B}_{*}-\Xi_B\right) \,,\qquad \Xi_B:=B^iD_iN+N^iD_iB_{*} \,.
\eeq
We see that $\Lag_3$ gives a non-trivial contribution to the canonical momenta conjugate to $B_{*}$ and $\gamma_{ij}$. However the fact that $\dot{B}_{*}$ appears only linearly still ensures the degeneracy.

\subsubsection{$\Lag_4$ term}

We work out $\Lag_4$ in two steps. The non-minimal coupling to the curvature scalar is straightforward to expand but it must be integrated by parts so as to remove second time derivatives. Thus we have
\beq\bal
\Lag_4^{(1)}&:=G_4R \\
&=G_4\left(K^{ij}K_{ij}-K^2+R^{(3)}+2\nabla_{\mu}(Kn^{\mu}-a^{\mu})\right) \\
&=G_4\left(K^{ij}K_{ij}-K^2+R^{(3)}\right)-2G_4'\nabla_{\mu}Y(Kn^{\mu}-a^{\mu})+{\rm t.d.} \\
&=G_4\left(K^{ij}K_{ij}-K^2+R^{(3)}\right)+2G_4'\bigg[2B^iB^jKK_{ij}+2B_{*}W_{*}K-2B^iF_iK \\
&\quad -2B^iD_iB_{*}K+\frac{1}{N}\,D^iND_iY\bigg]+{\rm t.d.}\,,
\eal\eeq
where $K:=\gamma^{ij}K_{ij}$, $R^{(3)}$ is the curvature scalar built out of $\gamma_{ij}$, and ``t.d.'' means total derivative. Next, the minimally covariantized GP term is
\beq\bal
\Lag_4^{(2)}&:=\widetilde{G}_4\Big[(\nabla_{\mu}B^{\mu})^2-\nabla_{\mu}B^{\nu}\nabla_{\nu}B^{\mu}\Big] \\
&=\widetilde{G}_4\bigg[-B_{*}^2(K^{ij}K_{ij}-K^2)+2B^iB^jK_i^{\phantom{i}k}K_{jk}-2B^iF^jK_{ij}+2B_{*}W_{*}K \\
&\quad -2D_iB^iW_{*}-2D_iB^iB_{*}K-4B_iD_jB_{*}K^{ij}+2B_{*}D_iB_jK^{ij}+2D_iB_{*}F^i \\
&\quad +2D^iB_{*}D_iB_{*}+(D_iB^i)^2-D_iB^jD_jB^i\bigg]\,.
\eal\eeq

Note that we have ``detuned'' the relative coefficients multiplying $\Lag_4^{(1)}$ and $\Lag_4^{(2)}$ so that we may understand later the role it plays in the coupled DHOST-GP system. When taken in isolation, however, we see that $\Lag_4$ contains
\beq
\Lag_4\supset 2\left(\widetilde{G}_4+2G_4'\right)B_{*}W_{*}K \,,
\eeq
which mixes the Proca field and metric velocities, and thus spoils the degeneracy unless we choose $\widetilde{G}_4=-2G_4'$, in agreement with \eqref{eq:GP terms def}.

\subsubsection{$\Lag_5$ term}

To expand $\Lag_5$ we consider the two contributions separately, again keeping the GP ``tuning'' of relative coefficients for later,
\beq\bal
\Lag_5^{(1)}&:=G_5(Y)G^{\mu\nu}\nabla_{\mu}B_{\nu} \,,\\
\Lag_5^{(2)}&:=\widetilde{G}_5(Y)\Big[(\nabla_{\mu}B^{\mu})^3-3(\nabla_{\rho}B^{\rho})\nabla_{\mu}B^{\nu}\nabla_{\nu}B^{\mu}+2\nabla_{\mu}B^{\nu}\nabla_{\nu}B^{\rho}\nabla_{\rho}B^{\mu}\Big] \,.
\eal\eeq
For the sake of brevity we will focus here on the kinetic terms, i.e.\ the terms which are at least quadratic in the velocities, delegating the full expressions to Appendix \ref{sec:appA}.

For the first contribution we need the components of the Einstein tensor in ADM variables,
\beq\bal
G^{00}&=-\frac{1}{2N^2}\left(K^{kl}K_{kl}-K^2-R^{(3)}\right) \,,\\
G^{0i}&=\frac{N^i}{2N^2}\left(K^{kl}K_{kl}-K^2-R^{(3)}\right)-\frac{1}{N}\,D_j\widetilde{K}^{ij} \,,\\
G^{ij}&=G^{(3)ij}-\frac{N^iN^j}{2N^2}\left(K^{kl}K_{kl}-K^2-R^{(3)}\right)+\frac{1}{N\sqrt{\gamma}}\,\partial_t\left(\sqrt{\gamma}\,\widetilde{K}^{ij}\right) \\
&\quad -\frac{1}{N}\left(D_iD_jN-\gamma^{ij}D^2N\right)+\frac{1}{N}\,D_k\left(2N^{(i}\widetilde{K}^{j)k}-N^k\widetilde{K}^{ij}\right) \\
&\quad +2\left(K^i_{\phantom{i}k}K^{jk}-KK^{ij}\right)-\frac{1}{2}\,\gamma^{ij}\left(K^{kl}K_{kl}-K^2\right) \,,
\eal\eeq
where $\widetilde{K}_{ij}:=K_{ij}-\gamma_{ij}K$ and $G^{(3)}_{ij}$ is the Einstein tensor built out of the 3-metric. Let us emphasize that the last result is only valid in three spatial dimensions. After collecting terms and integrating by parts we obtain
\beq
\Lag_{5,{\rm kin}}^{(1)}= \frac{1}{2N}\,G_5'\dot{Y}\left[B_{*}(K^{ij}K_{ij}-K^2)-2\widetilde{K}^{ij}D_iB_j\right] +{\rm t.d.}
\eeq
The result is proportional to $G_5'(Y)$ (the derivative of $G_5(Y)$ with respect to its argument), not surprisingly since $\Lag_5^{(1)}$ is a total derivative when $G_5$ is constant. Note that
\beq
\dot{Y}=2\left(-B_{*}\dot{B}_{*}+B^k\dot{B}_k-NB^kB^lK_{kl}-B^kB^lD_kN_l\right) \,.
\eeq

Expanding next $\Lag_5^{(2)}$ we eventually find
\beq\bal
\Lag_{5,{\rm kin}}^{(2)}&= \widetilde{G}_5\bigg\{\frac{3}{N}\,\dot{B}_{*}B_{*}\left[B_{*}(K^{ij}K_{ij}-K^2)-2\widetilde{K}^{ij}D_iB_j\right] \\
&\quad -B_{*}^3\left[K^3-3KK^{ij}K_{ij}+2K^i_{\phantom{i}j}K^j_{\phantom{j}k}K^k_{\phantom{k}i}\right]+6B_{*}B^iB^j\left[K^{kl}K_{ik}K_{jl}-KK_i^{\phantom{i}k}K_{jk}\right] \\
&\quad -\frac{3}{N}(K^{ij}K_{ij}-K^2)B_{*}^2\left[D_k(NB^k)+N^kD_kB_{*}\right] \\
&\quad +6(K_i^{\phantom{i}k}K_{jk}-KK_{ij})B_{*}\left[B_{*}D^iB^j-F^iB^j-2B^iD^jB_{*}\right] \\
&\quad -6B^iB^j\left[D^kB^lK_{ik}K_{jl}-D_lB^lK_i^{\phantom{i}k}K_{jk}\right] \\
&\quad +6F_i\left[\widetilde{K}^{ik}B_{*}D_kB_{*}+K^{kl}B_kD^iB_l-K^{ik}B_kD_lB^l\right]\bigg\} \,.
\eal\eeq
Comparing the two contributions we see that the offending terms proportional to $\dot{B}_{*}$ are indeed canceled upon choosing $\widetilde{G}_5=\frac{1}{3}\,G_5'$.

%%%%%%%%%%%%%%%%%%%%%%%%%%%%%%%%%%%%%%%%%%
\section{Constraint analysis}

In this section we collect the contributions to the GP terms in the Hamiltonian formalism and analyze the conditions for the Proca and DHOST constraints to be maintained once the two sectors are coupled through the metric tensor. We focus on each GP term independently, although in the end it will become clear that the results remain unchanged if one includes the whole Lagrangian. 

\subsection{$\Lag_3$ term}

We consider the addition to the gravitational action \eqref{eq:full quad dhost} the following GP vector matter term:
\beq\bal
S_{\rm m}&=\int d^4x\sqrt{-g}\bigg[G_2(Y,{\cal F},{\cal G})+G_3(Y)\nabla_{\mu}B^{\mu}\bigg] \\
&=\int dtd^3x N\sqrt{\gamma}\Big[2{\cal C}^{ij}_{\rm m}K_{ij}+2\overline{\cal C}^0W_{*}-{\cal U}_{\rm m}\Big] \,,
\eal\eeq
where
\beq\bal
{\cal C}^{ij}_{\rm m}&=-\frac{1}{2}\,G_3B_{*}\gamma^{ij} \,,\\
\overline{\cal C}^0&=-\frac{1}{2}\,G_3 \,,\\
{\cal U}_{\rm m}&=-G_2-G_3D_iB^i \,.
\eal\eeq
With some abuse of terminology we can think of ${\cal U}_{\rm m}$ as a potential term because it is independent of $\dot{B}_{*}$ and $K_{ij}$, however one should keep in mind that it does depend on $\dot{B}_i$.

The complete action $S_{\rm g}+S_{\rm m}$ is manifestly degenerate because the Hessian matrix is not affected by $S_{\rm m}$ as far as $\dot{B}_{*}$, $\dot{A}_{*}$ and $K_{ij}$ are concerned. Nevertheless the primary constraints are still affected by the linear terms (in the velocities) brought in by $\Lag_1$. In particular the Proca constraint is modified as follows:
\beq\begin{gathered}
q_{*}:=\frac{\partial\Lag}{\partial\dot{B}_{*}}=\frac{1}{N}\,\frac{\partial\Lag}{\partial W_{*}}=-\sqrt{\gamma}\,G_3 \\
\Rightarrow\qquad  \Lambda:=q_{*}+\sqrt{\gamma}\,G_3 \approx 0 \,.
\end{gathered}\eeq
To obtain the DHOST constraint we first compute the momenta
\beq\bal
\pi^{ij}&:=\frac{\partial\Lag}{\partial\dot{\gamma}_{ij}}=\frac{1}{2N}\,\frac{\partial\Lag}{\partial K_{ij}}=\sqrt{\gamma}\left[{\cal K}^{ij,kl}K_{kl}+{\cal B}^{ij}V_{*}+{\cal C}^{ij}_{\rm tot}\right] \,,\\
p_{*}&:=\frac{\partial\Lag}{\partial\dot{A}_{*}}=\frac{1}{N}\,\frac{\partial\Lag}{\partial V_{*}}=2\sqrt{\gamma}\left[{\cal A}V_{*}+{\cal B}^{ij}K_{ij}+{\cal C}^0\right] \,,
\eal\eeq
\beq
\Rightarrow\qquad \Psi':=p_{*}-2{\cal K}^{-1}_{ij,kl}\pi^{ij}{\cal B}^{kl}+2\sqrt{\gamma}\left({\cal K}^{-1}_{ij,kl}{\cal C}^{ij}_{\rm tot}{\cal B}^{kl}-{\cal C}^0\right) \approx 0 \,,
\eeq
where ${\cal C}^{ij}_{\rm tot}={\cal C}^{ij}+{\cal C}^{ij}_{\rm m}$ includes the contribution from the matter action. This is to be compared with the ``vacuum'' constraint
\beq
\Psi:=p_{*}-2{\cal K}^{-1}_{ij,kl}\pi^{ij}{\cal B}^{kl}+2\sqrt{\gamma}\left({\cal K}^{-1}_{ij,kl}{\cal C}^{ij}{\cal B}^{kl}-{\cal C}^0\right) \approx 0 \,,
\eeq
that one would have in the absence of matter. We conclude that there is no inconsistency at this stage: the $\Lag_3$ GP term maintains the primary constraints in the coupled GP-DHOST theory and is therefore safe with regards to the criterion (I) explained in the introduction.

The inconsistency of the model is manifested in the failure to generate the secondary constraints that $S_{\rm g}$ and $S_{\rm m}$ possess when taken in isolation. That is, the model fails criterion (II). This is because the primary constraints $\Psi'$ and $\Lambda$ do not Poisson-commute,
\beq\bal \label{eq:L3 secondary constraint}
\left\{\Psi',\Lambda \right\}&=2\sqrt{\gamma}\,{\cal K}^{-1}_{ij,kl}\left[\frac{\delta{\cal C}^{ij}_{\rm m}}{\delta B_{*}}+\frac{1}{\sqrt{\gamma}}\,\frac{\delta\Lambda}{\delta \gamma_{ij}}\right]{\cal B}^{kl} \\
&=2\sqrt{\gamma}\,{\cal K}^{-1}_{ij,kl}G_3'(Y)\left(B_{*}^2\gamma^{ij}-B^iB^j\right){\cal B}^{kl} \,.
\eal\eeq
Clearly $G_3'\neq 0$ since otherwise $\Lag_3$ is a total derivative and hence trivial. Thus the only way for the constraints to commute is that
\beq \label{eq:L3 secondary constraint2}
{\cal K}^{-1}_{ij,kl}\left(B_{*}^2\gamma^{ij}-B^iB^j\right){\cal B}^{kl}=0 \,.
\eeq
Recall that this condition should be understood as an identity valid for all field configurations. It implies a set of equations for the coefficients $A_I$. We find that \eqref{eq:L3 secondary constraint2} has a unique solution when complemented with the DHOST constraint \eqref{eq:dhost cond},
\beq \label{eq:L3 dhost sol}
A_1=-A_2=2F_X \,,\qquad A_3=A_4=A_5=0 \,,
\eeq
which is in fact the same solution that yields ${\cal B}^{ij}=0={\cal A}$. The vanishing of both ${\cal B}^{ij}$ and ${\cal A}$ is trivially a sufficient condition for both constraints to hold; what we have proved is that it is also a necessary condition.

With the result \eqref{eq:L3 dhost sol} for the functions $A_I$ the covariant DHOST action reduces to
\beq \label{eq:pure horndeski}
S_{\rm g}=\int d^4x\sqrt{-g}\Big[FR+2F_X\left(\nabla^{\mu}\nabla^{\nu}\phi\nabla_{\mu}\nabla_{\nu}\phi-(\Box\phi)^2\right)\Big] \,,
\eeq
which is nothing but the non-degenerate quadratic (in $\nabla^2\phi$) Horndeski Lagrangian.

\subsection{$\Lag_4$ term}

Next we consider adding to the DHOST action \eqref{eq:full quad dhost} the $\Lag_4$ GP term:
\beq\bal
S_{\rm m}&=\int d^4x\sqrt{-g}\bigg[G_2(Y,{\cal F},{\cal G})+G_4(Y)R+\widetilde{G}_4(Y)\Big((\nabla_{\mu}B^{\mu})^2-\nabla_{\mu}B^{\nu}\nabla_{\nu}B^{\mu}\Big)\bigg] \\
&=\int dtd^3x N\sqrt{\gamma}\Big[2\overline{\cal B}^{ij}W_{*}K_{ij}+{\cal K}^{ij,kl}_{\rm m}K_{ij}K_{kl}+2{\cal C}^{ij}_{\rm m}K_{ij}+2\overline{\cal C}^0W_{*}-{\cal U}_{\rm m}\Big] \,,
\eal\eeq
where
\beq\bal
\overline{\cal B}^{ij}&=(2G_4'+\widetilde{G}_4)B_{*}\gamma^{ij} \,,\\
{\cal K}^{ij,kl}_{\rm m}&=\left(G_4-\widetilde{G}_4B_{*}^2\right)\left(\gamma^{k(i}\gamma^{j)l}-\gamma^{ij}\gamma^{kl}\right)+2G_4'\left(\gamma^{ij}B^kB^l+\gamma^{kl}B^iB^j\right)+2\widetilde{G}_4B^{(i}\gamma^{j)(k}B^{l)} \,,\\
{\cal C}^{ij}_{\rm m}&=-2G_4'B^k(F_k+D_kB_{*})\gamma^{ij}-\widetilde{G}_4\left(B^{(i}F^{j)}+B_{*}D_kB^k\gamma^{ij}-B_{*}D^{(i}B^{j)}+B^{(i}D^{j)}B_{*}\right) \,,\\
\overline{\cal C}^0&=-\widetilde{G}_4D_iB^i \,,\\
{\cal U}_{\rm m}&=-G_2-G_4R^{(3)}-\frac{2G_4'}{N}\,D^iND_iY -\widetilde{G}_4\left[2D^iB_{*}(F_i+D_iB_{*})+(D_iB^i)^2-D_iB^jD_jB^i\right] \,.
\eal\eeq
As before, we are abusing the notation by including the velocities $\dot{B}_i$ (contained in the definition of $F_i$, see eq.\ \eqref{eq:def F_i}) into the above coefficient tensors. Now, however, it should be noted that $\dot{B}_i$ mixes with the extrinsic curvature, and this has important consequences as we explain next.

The critical question is whether we can find analogues of the DHOST and Proca primary constraints for this theory. To address this we compute the canonical momenta,
\beq\bal \label{eq:L4 momenta}
q_{*}&=2\sqrt{\gamma}\left[\overline{\cal B}^{ij}K_{ij}+\overline{\cal C}^0\right] \,,\\
p_{*}&=2\sqrt{\gamma}\left[{\cal A}V_{*}+{\cal B}^{ij}K_{ij}+{\cal C}^0\right] \,,\\
\pi^{ij}&=\sqrt{\gamma}\left[({\cal K}^{ij,kl}+{\cal K}^{ij,kl}_{\rm m})K_{kl}+{\cal B}^{ij}V_{*}+\overline{\cal B}^{ij}W_{*}+{\cal C}^{ij}_{\rm tot}\right] \,,\\
q^i&=\sqrt{\gamma}\left[2K_{jk}\,\frac{\partial{\cal C}^{jk}_{\rm m}}{\partial F_i}-\frac{\partial{\cal U}_{\rm m}}{\partial F_i}\right] \,,
\eal\eeq
where ${\cal C}^{ij}_{\rm tot}={\cal C}^{ij}+{\cal C}^{ij}_{\rm m}$, while $q^i:=\partial\Lag/\partial\dot{B}_i$ denotes the momentum conjugate to $B_i$. In order for the two constraints to exist the Hessian matrix must possess two independent null eigenvectors. We will demand that one of them be along the $\dot{B}_{*}$ direction---this is essentially what we mean by a GP theory, although it is in principle possible that the Proca constraint be realized in a more general way. This vector will be a null eigenvector if and only if $\overline{\cal B}^{ij}=0$ identically, and so we recover the usual relation $\widetilde{G}_4=-2G_4'$ of GP theory.

Investigating the existence of the DHOST constraint is complicated in this case because of the presence of $F_i$ on the right-hand side of the system \eqref{eq:L4 momenta}. As we assume that the GP sector has no further degeneracies beyond the one implied by the Proca constraint, the last equation in \eqref{eq:L4 momenta} can be used to express $F_i$ in terms of $K_{ij}$ and the canonical variables. There are two possibilities: (i) the relation between $F_i$ and $K_{ij}$ is linear, in which case this can substituted into the coefficient ${\cal C}^{ij}_{\rm m}$ in the third equation so as to obtain a linear system involving only the velocities $V_{*}$ and $K_{ij}$; (ii) the solution for $F_i$ depends non-linearly on $K_{ij}$, in which case the resulting system for $V_{*}$ and $K_{ij}$ will also be non-linear. Option (ii) is clearly inconsistent with the DHOST constraint, since the non-linear system thus obtained cannot be degenerate except in trivial cases. We will encounter the same situation when analyzing the $\Lag_5$ term in the next subsection, where we give further comments about this issue.

Focusing then on option (i), the most general way to achieve a linear relation between $F_i$ and $K_{ij}$ is by choosing the $G_2$ function as
\beq
G_2(Y,{\cal F},{\cal G})=g_2(Y){\cal F}+\widetilde{g}_2(Y){\cal G} \,,
\eeq
where $g_2(Y)$ and $\widetilde{g}_2(Y)$ are generic functions. This is to substituted into the last equation in \eqref{eq:L4 momenta}, which one then has to solve for $F_i$. Plugging the result into the coefficient ${\cal C}^{ij}_{\rm m}$ one finds
\beq
{\cal C}^{ij}_{\rm m}=g_4\left[KB^iB^j-B^{(i}K^{j)k}B_k-\gamma^{ij}\left(K^2B^kB_k-B^kB^lK_{kl}\right)\right]+{\cal D}^{ij} \,,
\eeq
where ${\cal D}^{ij}$ depends solely on the canonical variables (and not on $K_{ij}$) and $g_4:=\frac{8G_4^{\prime2}}{g_2-2\widetilde{g}_2Y}$ is a useful shorthand notation. Using this in \eqref{eq:L4 momenta} we arrive at the following reduced system:
\beq\bal
p_{*}&=2\sqrt{\gamma}\left[{\cal A}V_{*}+{\cal B}^{ij}K_{ij}+{\cal C}^0\right] \,,\\
\pi^{ij}&=\sqrt{\gamma}\left[{\cal K}^{ij,kl}_{\rm tot}K_{kl}+{\cal B}^{ij}V_{*}+{\cal D}^{ij}\right] \,,\\
\eal\eeq
where ${\cal K}^{ij,kl}_{\rm tot}:={\cal K}^{ij,kl}+{\cal K}^{ij,kl}_{\rm eff}$ and
\beq\bal \label{eq:Keff def}
{\cal K}^{ij,kl}_{\rm eff}&=\left(G_4+2G_4'B_{*}^2\right)\gamma^{k(i}\gamma^{j)l}-\left(G_4+2G_4'B_{*}^2+g_4(Y+B_{*}^2)\right)\gamma^{ij}\gamma^{kl} \\
&\quad +\left(2G_4'+g_4\right)\left(\gamma^{ij}B^kB^l+\gamma^{kl}B^iB^j\right)-\left(4G_4'+g_4\right)B^{(i}\gamma^{j)(k}B^{l)} \,.
\eal\eeq
The DHOST constraint will then be present if and only if
\beq \label{eq:L4 modified dhost cond}
{\cal A}-{\cal K}^{-1}_{{\rm tot}\,ij,kl}{\cal B}^{ij}{\cal B}^{kl}=0 \,.
\eeq
There are two ways for this relation to hold. First, we may choose to define the DHOST sector independently of the GP sector, so that we would have the usual constraint ${\cal A}={\cal K}^{-1}_{ij,kl}{\cal B}^{ij}{\cal B}^{kl}$. This would be in line with the treatment of the GP vector as a matter field which couples to the gravitational sector described by DHOST only through the metric, in the same way as any other matter field. The second way is to include the GP vector field in the very definition of the DHOST Lagrangian and impose the condition ${\cal A}={\cal K}^{-1}_{{\rm tot}\,ij,kl}{\cal B}^{ij}{\cal B}^{kl}$ as a constraint on the coefficient functions. This option would be akin to constructing a particular type of scalar-vector-tensor model from the bottom-up, and is therefore beyond our current scope exposed in the introduction.

Focusing then on the first possibility, we investigate if the equation
\beq \label{eq:L4 dhost constraint-1}
\left({\cal K}^{-1}_{ij,kl}-{\cal K}^{-1}_{{\rm tot}\,ij,kl}\right){\cal B}^{ij}{\cal B}^{kl}=0 \,,
\eeq
could hold as an identity. We first note that the matrix on the left-hand side can be written as
\beq
{\cal K}^{-1}_{ij,kl}-{\cal K}^{-1}_{{\rm tot}\,ij,kl}={\cal K}^{-1}_{{\rm tot}\,ij,mn}{\cal K}^{mn,pq}_{\rm eff}{\cal K}^{-1}_{pq,kl} \,.
\eeq
Inverting the matrix ${\cal K}_{\rm tot}$ requires some formidable amount of algebra, so for convenience we will expand perturbatively in the Proca field $B_{\mu}$, i.e.\
\beq \label{eq:Km expansion}
{\cal K}^{-1}_{ij,kl}-{\cal K}^{-1}_{{\rm tot}\,ij,kl}={\cal K}^{-1}_{ij,mn}{\cal K}_{\rm eff}^{mn,pq}{\cal K}^{-1}_{pq,kl}-{\cal K}^{-1}_{ij,mn}{\cal K}_{\rm eff}^{mn,pq}{\cal K}^{-1}_{pq,rs}{\cal K}_{\rm eff}^{rs,tu}{\cal K}^{-1}_{tu,kl}+{\cal O}({\cal K}_m^3) \,.
\eeq
Note that it does not matter at which order in $B_{\mu}$ the tensor ${\cal K}_{\rm eff}$ starts. Indeed from \eqref{eq:Keff def} we see that, regardless of the form of $G_4(Y)$ and $g_4(Y)$, each tensor structure in ${\cal K}_{\rm eff}$ starts at the same order in $B_{\mu}$. This may seem to require that $G_4(Y)$ be an analytic function of $Y$, however in reality all we demand is that there exists a field configuration for which an expansion in powers of ${\cal K}_{\rm eff}$ is admissible, as in eq.\ \eqref{eq:Km expansion}. For instance any $G_4(Y)$ admitting a Laurent series representation near $Y=0$ would give such consistent expansion.

Eq.\ \eqref{eq:L4 dhost constraint-1} together with the DHOST condition \eqref{eq:dhost cond} give two equations that must be satisfied identically. After replacing $A^iA_i=X+A_{*}^2$, $B^iB_i=Y+B_{*}^2$ and $A^iB_i=Z+A_{*}B_{*}$ (here $Z:=A^{\mu}B_{\mu}$), the equations can be expanded in powers of $A_{*}$ and $B_{*}$ so that the coefficient of each monomial must separately vanish. This yields a system of equations which, at leading order in ${\cal K}_{\rm eff}$, i.e.\ keeping only the first term on the RHS of \eqref{eq:Km expansion}, involves only the DHOST functions $A_I$ and $F$. We find that this system admits a single solution corresponding to ${\cal B}^{ij}=0$, which of course solves the degeneracy conditions not just to leading order in $B_{\mu}$ but in general.

In conclusion, the unique consistent solution to the degeneracy conditions \eqref{eq:L4 dhost constraint-1} and \eqref{eq:dhost cond} is the trivial one with ${\cal B}^{ij}=0$, which takes us again back to \eqref{eq:pure horndeski}, i.e.\ the standard Horndeski scalar-tensor theory.

\subsection{$\Lag_5$ term}

Focusing next on the $\Lag_5$ GP term we envisage the matter action
\beq\bal \label{eq:L5 adm}
S_{\rm m}&=\int d^4x\sqrt{-g}\bigg[G_2(Y,{\cal F},{\cal G})+G_5(Y)G^{\mu\nu}\nabla_{\mu}B_{\nu} \\
&\quad +\widetilde{G}_5(Y)\Big((\nabla_{\mu}B^{\mu})^3-3(\nabla_{\rho}B^{\rho})\nabla_{\mu}B^{\nu}\nabla_{\nu}B^{\mu}+2\nabla_{\mu}B^{\nu}\nabla_{\nu}B^{\rho}\nabla_{\rho}B^{\mu}\Big)\bigg] \\
&=\int dtd^3x N\sqrt{\gamma}\Big[2\overline{\cal D}^{ij,kl}W_{*}K_{ij}K_{kl}+2\overline{\cal B}^{ij}W_{*}K_{ij}+\frac{2}{3}\,{\cal J}^{ij,kl,mn}_{\rm m}K_{ij}K_{kl}K_{mn} \\
&\quad +{\cal K}^{ij,kl}_{\rm m}K_{ij}K_{kl}+2{\cal C}^{ij}_{\rm m}K_{ij}+2\overline{\cal C}^0W_{*}-{\cal U}_{\rm m}\Big] \,,
\eal\eeq
where
\beq\bal
\overline{\cal D}^{ij,kl}&=-\frac{1}{2}(G_5'-3\widetilde{G}_5)B_{*}^2\left(\gamma^{i(k}\gamma^{l)k}-\gamma^{ij}\gamma^{kl}\right) \,,\\
\overline{\cal B}^{ij}&=(G_5'-3\widetilde{G}_5)B_{*}\left(D^{(i}B^{j)}-\gamma^{ij}D_kB^k\right) \,,\\
{\cal J}^{ij,kl,mn}_{\rm m}&=-\frac{G_5'}{2}\left[B^iB^j\gamma^{k(m}\gamma^{n)l}-B^iB^j\gamma^{kl}\gamma^{mn}+(2\,{\rm perm.})\right] \\
&\quad +3\widetilde{G}_5\left[B^iB^k\gamma^{j(m}\gamma^{n)l}-B^iB^k\gamma^{jl}\gamma^{mn}+(2\,{\rm perm.})\right] \\
&\quad -\frac{3\widetilde{G}_5}{2}\,B_{*}\Big[\gamma^{ij}\gamma^{kl}\gamma^{mn}-\left(\gamma^{ij}\gamma^{k(m}\gamma^{n)l}+(2\,{\rm perm.})\right) \\
&\quad +\left(\gamma^{j(k}\gamma^{l)(m}\gamma^{n)i}+\gamma^{i(k}\gamma^{l)(m}\gamma^{n)j}\right)\Big] \,,\\
{\cal K}^{ij,kl}_{\rm m}&=G_5'\bigg[B_{*}\left(B^mF_m+B^mD_mB_{*}+\frac{1}{2N}\,N^mD_mY\right)\left(\gamma^{i(k}\gamma^{l)k}-\gamma^{ij}\gamma^{kl}\right) \\
&\quad +\left(B^iB^jD^{(k}B^{l)}+B^kB^lD^{(i}B^{j)}\right)-D_mB^m\left(B^iB^j\gamma^{kl}+B^kB^l\gamma^{ij}\right)\bigg] \\
&\quad +6\widetilde{G}_5\bigg[-\frac{1}{2N}\,B_{*}^2\left(D_m(NB^m)+N^mD_mB_{*}\right)\left(\gamma^{i(k}\gamma^{l)k}-\gamma^{ij}\gamma^{kl}\right) \\
&\quad +B_{*}\left((B_{*}D^{i}B^{(k}-B^{i}F^{(k}-2B^{i}D^{(k}B_{*})\gamma^{l)j}+(1\,{\rm perm.})\right) \\
&\quad -B_{*}\left((B_{*}D^{(i}B^{j)}-B^{(i}F^{j)}-2B^{(i}D^{j)}B_{*})\gamma^{kl}+(1\,{\rm perm.})\right) \\
&\quad -\left(B^{(i}D^{j)}B^{(k}B^{l)}-D_mB^mB^{(i}\gamma^{j)(k}B^{l)}\right)\bigg] \,.
\eal\eeq
Once again we abuse the notation to include terms involving $\dot{B}_i$ in these coefficient tensors. The remaining coefficients entering in \eqref{eq:L5 adm}, the ones at most linear in the velocity variables, are provided in full in Appendix \ref{sec:appA}.

The relevant set of canonical momenta is given by
\beq\bal \label{eq:L5 momenta}
q_{*}&=2\sqrt{\gamma}\left[\overline{D}^{ij,kl}K_{ij}K_{kl}+\overline{\cal B}^{ij}K_{ij}+\overline{\cal C}^0\right] \,,\\
p_{*}&=2\sqrt{\gamma}\left[{\cal A}V_{*}+{\cal B}^{ij}K_{ij}+{\cal C}^0\right] \,,\\
\pi^{ij}&=\sqrt{\gamma}\left[{\cal J}^{ij,kl,mn}_{\rm m}K_{kl}K_{mn}+2\overline{D}^{ij,kl}W_{*}K_{kl}+({\cal K}^{ij,kl}+{\cal K}^{ij,kl}_{\rm m})K_{kl}+{\cal B}^{ij}V_{*}+\overline{\cal B}^{ij}W_{*}+{\cal C}^{ij}_{\rm tot}\right] \,,\\
\eal\eeq
where ${\cal C}^{ij}_{\rm tot}={\cal C}^{ij}+{\cal C}^{ij}_{\rm m}$. It is clear that in the absence of the GP tuning the standard Proca constraint fails to be realized, i.e.\ the Hessian matrix does not have a null eigenvector along the $\dot{B}_{*}$ direction. As before, we will insist that this eigenvector be present while keeping in mind that other options may in principle be available. Therefore at this stage we set $\widetilde{G}_5=\frac{1}{3}\,G_5'$, so that in particular $\overline{D}^{ij,kl}=0=\overline{\cal B}^{ij}$.

To the system \eqref{eq:L5 momenta} one must also add the relation for the canonical momentum conjugate to $\dot{B}_i$, which as before is to be solved for $F_i$ in terms of $K_{ij}$. This relation is now unavoidably non-linear because $F_i$ also enters in the tensor ${\cal K}^{ij,kl}_{\rm m}$. In addition, the system also involves terms quadratic in $K_{ij}$ because of the presence of the tensor ${\cal J}^{ij,kl,mn}_{\rm m}$ (which is non-zero since $G_5'\neq0$, otherwise $\Lag_5$ is a total derivative). Thus, the novelty brought in by the $\Lag_5$ GP term is that the relation between $K_{ij},V_{*}$ and the canonical variables is necessarily non-linear. Such system can only be degenerate in a trivial manner, i.e.\ if the coefficients are such that one of the variables disappears from the system. In this case, for $V_{*}$ to drop out, we must have ${\cal A}=0={\cal B}^{ij}$. The conclusion is that the $\Lag_5$ term of GP does not admit a consistent coupling to DHOST except in the non-degenerate case of Horndeski theory.

%%%%%%%%%%%%%%%%%%%%%%%%%%%%%%%%%%%%%%%%%%
\section{Discussion}

We have demonstrated that generalized Proca fields described by GP theory do not allow for a consistent coupling to a gravitational sector given by the DHOST class of models. Although our analysis considered the individual GP Lagrangians separately, it is clear in hindsight that none of the results would change if we were to envisage the complete model: the $\Lag_5$ GP term immediately spoils the DHOST degeneracy because of the cubic operators in the extrinsic curvature, while the $\Lag_4$ also fails the degeneracy test irrespective of $\Lag_3$. The exceptions that bypass our no-go result are rather trivial, at least from the perspective of the constraint structure: either the DHOST sector must reduce to the standard, non-degenerate Horndeski theory, or the GP sector must reduce to the $\Lag_2$ term which is independent of the Christoffel connection.

It is important to emphasize the relation between having the correct number of constraints and the consistency of the theory. The appearance of an additional degree of freedom in the DHOST sector as a consequence of the coupling with GP theory is expected to be associated with a ghost instability. This follows from the Ostrogradsky theorem, since the DHOST equations of motion are higher than second order and, when the constraint is thwarted, there is no degeneracy responsible for reducing the number of pieces of initial data. Because of this, the Hamiltonian in this situation is unbounded from below and an instability will be present. As usual, this instability may be non-linear, i.e.\ the ghost mode need not appear as a linear perturbation on every background field configuration, but it will necessarily manifest itself around some backgrounds or at the non-linear level, as it occurs with the Boulware-Deser ghost in massive gravity \cite{Boulware:1973my}.

We stressed in the introduction that our set-up relies on various assumptions which we think worth to reiterate. The GP-DHOST system we studied is not the most general one. The analysis of the full model including all known operators would be a straightforward extension of our work and we expect our main conclusions to remain unchanged. Indeed, the additional terms of the GP class that we have omitted contain operators that are cubic and quartic in powers of $\nabla_{\mu}B_{\nu}$, hence they are likely to lead to the same issues as the $\Lag_5$ GP term. More crucial was the assumed prescription for coupling the GP and DHOST sectors. The premise was that there exists a Jordan frame such that all matter fields experience gravity through the same metric tensor and that our Proca field follows suit. Relaxing this assumption would be tantamount to constructing a scalar-vector-tensor type of theory in which all three fields interact in a non-trivial way. It would be interesting to address this problem within the context of degenerate theories (see e.g.\ \cite{Charmousis:2012dw,Heisenberg:2018acv,Heisenberg:2018mxx,Kase:2018nwt,Ikeda:2019okp} for some recent related work).

Finally, an additional assumption was made in the analysis of the primary constraints, where we demanded that the Proca constraint had to match that of GP theory, that is with a Hessian null eigenvector that is such that the time component of the vector field is rendered non-dynamical (in some local frame). It would be intriguing to explore if this hypothesis might be dropped in order for the Proca and DHOST constraints to be realized in a way that would mix the canonical momenta associated to the vector and scalar fields. We remark that a related generalization of the Proca constraint has been studied recently in \cite{deRham:2020yet} in the context of pure vector-tensor theories. We plan to revisit these questions in future work.

\begin{acknowledgments}

The author thanks C\'edric Deffayet for a discussion and comments. This research was funded by the European Union's Horizon 2020 Research Council grant 724659 MassiveCosmo ERC-2016-COG.

\end{acknowledgments}

%%%%%%%%%%%%%%%%%%%%%%%%%%%%%%%%%%%%%%%%%%%
%%%%%%%%%%%%%%%%%%%%%%%%%%%%%%%%%%%%%%%%%%%

\appendix

\section{Full results of the ADM decomposition} \label{sec:appA}

In the main text we omitted the explicit results for some of the coefficient tensors appearing in the $3+1$-decomposed DHOST and $\Lag_5$ GP term Lagrangians. Those terms are not important for the constraint analysis, but we provide the full expressions in this Appendix for the sake of completeness.

The tensors ${\cal C}^{ij}$, ${\cal C}^0$ and ${\cal U}$ that enter in the DHOST action are
\beq\bal
{\cal C}^{ij}&=-\left(2F_XA^kD_kA_{*}+F_{\phi}A_{*}+\frac{1}{2}\,QA_{*}\right)\gamma^{ij}+A_{*}\bigg[-A_1D^{(i}A^{j)}-A_2D_KA^k\gamma^{ij} \\
&\quad +\frac{A_3}{2}\left(A^iA^jD_kA^k-A^kA^lD_kA_l\gamma^{ij}\right)+A_5A^kA^lD_kA_lA^iA^j\bigg] \,,\\
{\cal C}^0&= -\frac{1}{2}\,Q-\left(A_2-\frac{A_3}{2}\,A_{*}^2\right)D_iA^i-\left(\frac{A_3}{2}-A_5A_{*}^2\right)A^iA^jD_iA_j \\
&\quad +\left(A_3+A_4-2A_5A_{*}^2\right)A^iA_{*}D_iA_{*} \,,\\
{\cal U}&=-R^{(3)}+2D_i\left(F_XD^iX+F_{\phi}A^i\right)-P-QD_iA^i-\Big(A_1D^iA^jD_iA_j+A_2(D_iA^i)^2 \\
&\quad +A_3A^iA^jD_iA_jD_kA^k+A_4A^iA^jD_iA^kD_jA_k+A_5A^iA^jA^kA^lD_iA_jD_kA_l\Big) \\
&\quad +(2A_1-A_4A_{*}^2)D^iA_{*}D_iA_{*}+(A_4-4A_5A_{*}^2)A^iA^jD_iA_{*}D_jA_{*} \\
&\quad +2A_{*}D^iA_{*}\left(A_3A_iD)jA^j+A_4A^jD_jA_i+2A_5A_iA^jA^kD_jA_k\right) \,.
\eal\eeq

The tensors ${\cal C}^{ij}_{\rm m}$, $\overline{\cal C}^0$ and ${\cal U}_{\rm m}$ that enter in the $\Lag_5$ GP term are
\beq\bal
{\cal C}^{ij}_{\rm m}&=\frac{3\widetilde{G}_5}{N}\bigg[(NF_k-N^lb_{kl})\left(B_{*}D^{(i}B_{*}\gamma^{j)k}-B_{*}D^kB_{*}\gamma^{ij}+D^kB^{(i}B^{j)}-D_mB^mB^{(i}\gamma^{j)k}\right) \\
&\quad -\frac{B_{*}}{2}\Big(2ND^kB_{*}D_kB_{*}+N(D_lB^k)^2+2N^kD_kB_{*}D_lB^l+2B^kD_kND_lB^l \\
&\quad -2N^kD_kB^lD_lB_{*}+2N^kD^lB_kD_lB_{*}-ND_kB^lD_lB^k\Big)\gamma^{ij} \\
&\quad +NB_{*}\Big(D^iB_{*}D^jB_{*}-D^kB^{(i}D^{j)}B_k+D^{(i}B^{j)}D_kB^k\Big)+B_{*}N^k\Big(D_kB_{*}D^{(i}B^{j)} \\
&\quad -D_kB^{(i}D^{j)}B_{*}+D^{(i}B_kD^{j)}B_{*}\Big)+N^kB^{(i}\Big(D_lB^{j)}b^l_{\phantom{l}k}-b^{j)}_{\phantom{j)}k}D_lB^l\Big) \\
&\quad +NB^{(i}\Big(D^{j)}B_kD^kB_{*}+D_kB^{j)}D^kB_{*}-2D^{j)}B_{*}D_lB^l\Big)+B_{*}B^kD_kND^{(i}B^{j)}\bigg] \,,\\
\overline{\cal C}^0&= \frac{3\widetilde{G}_5}{2}\left[D_iB^jD_jB^i-(D_iB^i)^2\right] \,,\\
{\cal U}_{\rm m}&= -G_2-\frac{3\widetilde{G}_5}{N}\bigg[2(NF_i-N^jb_{ij})\Big(D^iB_{*}D_kB^k-D_kB_{*}D^iB^k\Big)+\frac{N}{3}\Big((D_iB^i)^3 \\
&\quad -3D_iB^iD_jB^kD_kB^j+2D_iB^jD_jB^kD_kB^i\Big)+2N\Big(D_iB^iD^jB_{*}D_jB_{*}-D^iB^jD_iB_{*}D_jB_{*}\Big) \\
&\quad +2N^i\Big(D_iB^jD_jB^kD_kB_{*}-D_jB^kD^jB_iD_kB_{*}-b_{ij}D^jB_{*}D_kB^k\Big)\bigg] \,.
\eal\eeq

\section{Inverse of DHOST kinetic tensor} \label{sec:appB}

The metric kinetic tensor that appears in the Hamiltonian analysis of DHOST has the following structure (see eq.\ \eqref{eq:full adm dhost}):
\beq\bal
{\cal K}^{ij,kl}&=a\gamma^{i(k}\gamma^{l)j}+b\gamma^{ij}\gamma^{kl}+c\left(\gamma^{ij}A^kA^l+\gamma^{kl}A^iA^j\right) +d\left(A^iA^{(k}\gamma^{l)j}+A^jA^{(k}\gamma^{l)i}\right)+eA^iA^jA^kA^l \,.
\eal\eeq
We wish to find the inverse tensor such that
\beq
{\cal K}^{ij,mn}{\cal K}^{-1}_{mn,kl}=\delta^i_{(k}\delta^j_{l)} \,.
\eeq
This is achieved by taking the ansatz
\beq\bal
{\cal K}^{-1}_{ij,kl}&=a'\gamma_{i(k}\gamma_{l)j}+b'\gamma_{ij}\gamma_{kl}+c'\left(\gamma_{ij}A_kA_l+\gamma_{kl}A_iA_j\right) +d'\left(A_iA_{(k}\gamma_{l)j}+A_jA_{(k}\gamma_{l)i}\right)+e'A_iA_jA_kA_l \,,
\eal\eeq
and a straightforward calculation yields
\beq\bal
a'&=\frac{1}{a} \,,\\
b'&=\frac{1}{a}\,\frac{x \left(c^2 x-b (2 d+e x)\right)-a b}{a^2+3 a b+a x (2 (c+d)+e x)+2 x \left(2 b d+b e x-c^2x\right)} \,,\\
c'&=\frac{1}{a}\,\frac{-a c+2 b d+b e x-c^2x}{a^2+3 a b+a x (2 (c+d)+e x)+2 x \left(2 b d+b e x-c^2x\right)} \,,\\
d'&=-\frac{1}{a}\,\frac{d}{a+d x} \,,\\
e'&=\frac{1}{a}\,\frac{-a^2 e+a \left(-3 b e+3 c^2+4 c d+2 d^2+d e x\right)+d \left(2 b d+b e x-c^2x\right)}{(a+d x) \left(a^2+3 a b+a x (2 (c+d)+e x)+2 x \left(2 b d+b e x-c^2 x\right)\right)} \,,
\eal\eeq
with $x:=A^iA_i$.

%%%%%%%%%%%%%%%%%%%%%%%%%%%%%%%%%%%%%%%%%%%
%%%%%%%%%%%%%%%%%%%%%%%%%%%%%%%%%%%%%%%%%%%

\bibliographystyle{apsrev4-1}
\bibliography{GP-DHOSTBiblio}

\end{document}